\documentclass[sigconf]{acmart}

\usepackage{algorithm2e}
\usepackage{booktabs}
\usepackage{pgfplotstable}
\usepackage{subcaption}
\usepackage{mathdots}
\usepackage{dblfloatfix}

\DeclareMathOperator{\OP}{OP}

\def\BibTeX{{\rm B\kern-.05em{\sc i\kern-.025em b}\kern-.08emT\kern-.1667em\lower.7ex\hbox{E}\kern-.125emX}}

\copyrightyear{2018}
\acmYear{2018}
\setcopyright{acmlicensed}
\acmConference[Woodstock '18]{Woodstock '18: ACM Symposium on Neural Gaze Detection}{June 03--05, 2018}{Woodstock, NY}
\acmBooktitle{Woodstock '18: ACM Symposium on Neural Gaze Detection, June 03--05, 2018, Woodstock, NY}
\acmPrice{15.00}
\acmDOI{10.1145/1122445.1122456}
\acmISBN{978-1-4503-9999-9/18/06}

\begin{document}

%
\title{Cross-Platform Performance Portability Using Highly Parametrized SYCL Kernels}

%
\newcommand{\codeplayaffiliation}{%
  \affiliation{%
    \institution{Codeplay Software Ltd.}
  }
}
\author{John Lawson}
\email{john@codeplay.com}
\codeplayaffiliation{}
\author{Mehdi Goli}
\email{mehdi.goli@codeplay.com}
\codeplayaffiliation{}
\author{Duncan McBain}
\email{duncan@codeplay.com}
\codeplayaffiliation{}
\author{Daniel Soutar}
\email{daniel.soutar@codeplay.com}
\codeplayaffiliation{}
\author{Louis Sugy}
\email{louis.sugy@codeplay.com}
\codeplayaffiliation{}

%
\renewcommand{\shortauthors}{Lawson, Goli, et al.}

%
\begin{abstract}

  Over recent years heterogeneous systems have become more prevalent across HPC
  systems, with over 100 supercomputers in the TOP500 incorporating GPUs or
  other accelerators. These hardware platforms have different performance
  characteristics and optimization requirements. In order to make the most of
  multiple accelerators a developer has to provide implementations of their
  algorithms tuned for each device. Hardware vendors provide libraries targeting
  their devices specifically, which provide good performance but frequently have
  different API designs, hampering portability.

  The SYCL programming model allows users to write heterogeneous programs using
  completely standard C++, and so developers have access to the power of C++
  templates when developing compute kernels. In this paper we show that by
  writing highly parameterized kernels for matrix multiplies and convolutions we
  achieve performance competitive with vendor implementations across different
  architectures. Furthermore, tuning for new devices amounts to choosing the
  combinations of kernel parameters that perform best on the hardware.

\end{abstract}

%
%

\ccsdesc[500]{Mathematics of computing~Mathematical software performance}
\ccsdesc[300]{Computing methodologies~Massively parallel algorithms}
\ccsdesc[300]{Computing methodologies~Parallel programming languages}
\ccsdesc[100]{Computing methodologies~Neural networks}
\ccsdesc[100]{Computing methodologies~Machine learning algorithms}

%
\keywords{SYCL, OpenCL, parallel computing, portability, GPGPU, performance}

%

%
\maketitle

\section{Introduction}

The power and speed of modern GPUs coupled with their increased programmability
through programming models such as SYCL~\cite{sycl}, OpenCL~\cite{opencl} and
CUDA~\cite{cuda} have enabled many applications to achieve previously
insurmountable performance. This has been one of the main driving factors behind the
rise in machine learning since 2012.  In the TOP500~\cite{top500}, 3 of the top
5 supercomputers in the world in terms of pure performance utilize GPUs, as do
the top 4 supercomputers in the Green500~\cite{green500}, where flops per watt
are considered.

Many hardware vendors provide highly specialized libraries for their platforms
to extract optimal performance from their hardware. As more companies and
computing centers move to support larger heterogeneous systems with more varied
hardware this means users have to write applications which handle multiple
different libraries for each available accelerator. This adds complexity both
for developers and system maintainers.

\section{Related Works}\label{sec:relatedworks}

Highly specialized libraries---such as ARM Compute Library~\cite{acl},
cuBLAS~\cite{nvidia2008cublas}, cuDNN~\cite{chetlur2014cudnn},
CUTLASS~\cite{kerrcutlass}, MIOpen~\cite{miopen}, Intel
MKL~\cite{wang2014intel} and MKL-DNN~\cite{mkldnn}---provide optimal performance
on the targeted hardware, however typically each of these libraries are
restricted to the vendor's hardware.

A common way of achieving performance across various platforms is to specialize
certain routines in a library or framework targeting a particular architecture.
There are various high performance vendor-specified libraries 
tuned for particular architectures.

For ARM devices, ARM Compute Library~\cite{acl} provides a set of low-level,
optimized primitives for machine-learning and linear algebra.  While for NVIDIA
devices, cuBLAS~\cite{nvidia2008cublas} and cuDNN~\cite{chetlur2014cudnn}
provide highly optimized implementations of standard linear algebra subroutines
and deep neural network routines, respectively.  NVIDIA also provides a higher
level CUTLASS~\cite{kerrcutlass} which takes advantage of CUDA C++ templates
meta programming  to deliver a high-performance GEMM computations allowing
developers to specialize their matrix multiplies inside their application.  

Similarly AMD provides MIOpen~\cite{miopen}, a deep learning acceleration
framework developed to deliver highly optimized DNN and GEMM routines for AMD
GPUs. For their multi-core and many-core systems, Intel provides
MKL~\cite{wang2014intel} for linear algebra routines, and MKL-DNN~\cite{mkldnn}
as an open source, high performance library for accelerating deep learning
frameworks.

Each of these libraries are optimized specifically for their target devices,
however do not provide performance portability. The limitations of these
libraries highlight two main key factors in providing this: the parallel
programming framework used to develop the library and the performance metrics of
devices.

\subsection{Parallel programming frameworks}

There are many different frameworks available to provide access to hardware
parallelism, with different approaches and programming styles. Most are based on
C or C++, with variations on the language to allow a developer to
specify how tasks can be parallelized.

The Intel Thread Building Blocks (TBB) library~\cite{pheatt2008intel} is a C++
template library for parallel programming on multi-core processors.  TBB only
provides parallelism on CPUs so applications developed on top of TBB are not
portable to GPU or accelerated platforms. 

On the other hand, CUDA~\cite{cuda} is used by NVIDIA as a low level framework
to deliver performance on NVIDA GPUs and accelerators, but not CPUs. Although
CUDA C++ supports template metaprogramming features which is necessary for
performance portability, the lack of CUDA support on non-NVIDIA architecture
prevents more widespread adoption.

A cross-platform framework like OpenCL~\cite{opencl} can be supported by various
vendors and platforms, however the low level interface of OpenCL hampers
development of easily tunable kernels that enable performance portability.
Although OpenCL 2.1 supports template metaprogramming within kernels, the kernel
interface itself cannot be templated, which limits the use of this feature.

SYCL~\cite{sycl, reyes2015sycl} is an open standard maintained by the Khronos
Group, a collaboration of companies, universities and individuals dedicated to
developing open standards, known for maintaining OpenCL, Vulkan, OpenGL and many
others alongside SYCL\@.  The standard is a royalty-free, cross-platform
abstraction layer that enables developers to utilize heterogeneous systems with
varied hardware while writing completely standard C++.

SYCL provides a highly parallel programming model designed to abstract away much
of the complexity of traditional parallel programming models like OpenCL\@.
Using SYCL instead of a lower level interface like OpenCL allows the developer
to write code using modern C++ language features, even within their accelerated
kernels. This includes the use of templates and specializations, as well as the
metaprogramming that these enable, instead of the heavy use of the preprocessor
typical in OpenCL applications to provide specializations within kernels for
data types, tile sizes and other constants.  By allowing SYCL kernels to fully
utilize C++, a developer can instead use templates to provide these compile time
constants, as well as easily create many specialized kernels with very little
additional code.

Combining the portability of OpenCL with modern C++ template metaprogramming
makes SYCL a strong parallel programming framework for performance portability
across different platforms. 

ComputeCpp~\cite{comutecpp}, developed by Codeplay Software, is currently the
only fully conformant implementation of the SYCL standard, supporting devices
from AMD, Intel, NVIDIA and others.  Several open-source libraries are
compatible with ComputeCpp, including SYCL-DNN~\cite{sycldnn},
VisionCpp~\cite{goli2016visioncpp,visioncpp}, SYCL-PRNG~\cite{syclprng}, as well
as contributed parts of Eigen~\cite{eigenweb,goli2017sycl} and
TensorFlow~\cite{tensorflow2015-whitepaper, goli2017accelerated,
goli2018tensorflow}.

\subsection{Performance Metrics}\label{sec:perfmetrics}

Tuning the performance of accelerated compute kernels is not a trivial task
requiring expertise in both parallel algorithms and hardware architectures.
Having said that, many hardware accelerators share common factors affecting
performance, allowing common algorithmic techniques for improving performance
across devices. 

We have studied common hardware-related optimization improvements on the
following devices: Intel CPUs and GPUs, ARM Mali GPUs, AMD GPUs and the Renesas
V3H and V3M accelerators. These metrics are used to develop parametric
convolution and matrix multiply algorithms, two of the most computationally
intense operations used in neural networks.

\subsubsection{Thread reusability}

The number of individual compute units available in hardware varies
significantly, from 2 in embedded systems to 84 in desktop GPUs e.g NVIDIA Tesla
series. Depending on the number of compute units available the size of workload
per thread can be parametrized in order to achieve maximal device occupancy.
However, there is a trade off between increasing the number of work-groups and
ensuring a sufficient workload per thread.  For a given problem, creating more
work groups or more threads will naturally decrease the amount of computation
that any one thread will have to do.

\subsubsection{Memory transactions}

In most accelerators, when a set of threads within a work-group loads data, an
entire block of data called a cache-line is fetched from memory.  This is based
on the assumption that threads within a work-group access contiguous elements of
data specified in the fetched block.
Assuming the threads within a work-group access the same address, the
permutation, masking or prediction has no effect on the loaded block of data,
which can lead to more optimizations in the compiler. Moreover loading a block
of data will reduce the number of memory transactions, improving
performance~\cite{nvidia2019PG}.

\subsubsection{Data reusability}

Memory accesses tend to be the bottleneck in GPU programming, despite
graphics memory typically being much faster than standard DDR memory. As a
result of this, many optimizations to GPU kernels involve maximizing data reuse
to minimize the amount of data that needs to be fetched from memory. Hardware
devices provide a memory hierarchy in order to reduce the memory access
bottleneck which is categorized into global, local, and private memory regions
in OpenCL terms, each of which can be controlled by programmers.

Common techniques to maximize data reusability include utilizing local memory as
a programmable cache to allow a work group to reuse data loaded from global
memory, and through using blocks of private memory to allow single threads to
reuse memory.  Private memory usually maps to hardware registers and so can be
accessed incredibly quickly. Provided there are sufficient registers, the more
we can re-use data located in registers, the better the performance.

Local memory access speeds are usually equivalent to caches.  Nowadays many
hardware vendors (e.g.\ Intel and ARM) provide large caches in front of their
global memory and so can achieve similar performance to using
programmer-controlled local memory when accessing data in a coalesced manner.
Therefore, some embedded devices like ARM's Mali G-71 GPU do not dedicate any
memory for local memory, instead relying on the cache and global memory. For
such devices using local memory can be costly and negatively affect performance.

\begin{table}
  \caption{Performance metrics of various compute devices, as discussed in
  Section~\ref{sec:perfmetrics}.}%
  \label{tab:perfmetric}
  \begin{tabular}{lrrr}
    \toprule
    \textbf{Device name} &
    \parbox{1.2cm}{\raggedleft{} \textbf{Cache line}} &
    \parbox{1.3cm}{\raggedleft{} \textbf{Local memory}} &
    \parbox{1.3cm}{\raggedleft{} \textbf{Compute units}} \\
    \midrule
    Intel Core i7-6700K CPU &  64 bytes  &    None  &  8 \\
    Intel Core i7-6700K GPU &  64 bytes  &  64 KiB  & 24 \\
    ARM Mali G71 GPU        &  64 bytes  &    None  &  8 \\
    Renesas V3M             & 128 bytes  & 447 KiB  &  2 \\
    Renesas V3H             & 128 bytes  & 409 KiB  &  5 \\
    AMD R9 Nano             & 128 bytes  &  32 KiB  & 64 \\
    \bottomrule
  \end{tabular}
\end{table}

\subsubsection{Vectorization}

Many GPU architectures have memory controllers which are designed to load and
store multiple elements at once, as typical graphics workloads involve 4-element
vectors. To get the best performance from these load-store units, computations
should make use of vector loads and stores. Another benefit of using vectors in
a GPU kernel even if the hardware does not support vector computations is that
each vector element can be computed separately, and so there is increased
instruction level parallelism.

Table~\ref{tab:perfmetric} summarizes the hardware features required for
abstracting performance metrics for convolution and matrix multiply algorithms.

\subsection{Research Niche}

In this paper, we have designed and developed parametric algorithms for matrix
multiplication and convolution---two of the most computationally intense
operations in neural networks---which can be tuned according to the performance
metrics of various devices.  Written on top of SYCL, these implementations
abstract common performance metrics, enabling performance portability across
different platforms.

\section{SYCL-BLAS}\label{sec:syclblas}

SYCL-BLAS~\cite{syclblas} is an open source implementation of netlib
BLAS~\cite{blasgemm} which enables performance portability across a range of
accelerators.
SYCL-BLAS uses an expression tree design to deliver BLAS co-routines; most of
the BLAS Level 1 and BLAS Level 2 co-routines are memory-bound operations so
using such an expression tree based approach allows multiple operations to be
fused into a single compute kernel with a higher computational complexity.
Increasing the computational intensity of memory-bound applications can
significantly increase the performance by reducing the number of accesses to the
device's global memory~\cite{aliaga2017sycl}.

The general matrix multiply operation (GEMM) is the heart of BLAS routines, used
in a wide range of scientific domains from physics to machine learning.  It is
one of the computationally expensive operation in neural networks, and so
optimizing GEMM can improve a DNN's performance.

\subsection{GEMM}\label{sec:gemm}

The General Matrix-Matrix product is a BLAS operation defined as:
\[
  C = \alpha\times \OP_{a}(A)\times \OP_{b}(B) + \beta\times C,
\]
where $A$, $B$ and $C$ are column-major matrices,
and $\OP_{a}$ and $\OP_{b}$ are either identity or transpose operators
(or conjugate transpose for complex data types).

We refer to $M$ and $N$ as the number of rows and columns of $C$, respectively,
and $K$ as the contracting dimension of $A$ and $B$. A naive parallelization
approach on massively parallel architectures is to assign one value of the
output $C$ per thread to accumulate the dot product of the $i$-th row of
$\OP_a(A)$ with the $j$-th column of $\OP_b(B)$ in a local register $r$, and
update the result: $C(i, j) = \alpha r + \beta C(i, j)$.  However,
this approach is highly memory-bounded as each thread is required to load
$2\times k$ data elements while performing roughly $2\times k$ floating point
operations.

A standard approach used to improve performance of dense linear algebra
operations on modern hardware is to reduce the memory latency by
increasing data reusability through a tiling technique~\cite{gustavson1998recursive,nath2010improved}.

\subsubsection{Blocked GEMM}

Given two matrices $A_{M \times K}$ and $B_{K \times N}$ and partitions
\begin{equation}
  \begin{split}
    0 & = M_0 < \dotsb < M_m = M \in [0, M]  \\
    0 & = K_0 < \dotsb < K_k = K \in [0, K]  \\
    0 & = N_0 < \dotsb < N_n = N \in [0, N]
  \end{split}
\end{equation}
then the following holds:
\begin{multline*}
  (A\times B)(M_{i-1}:M_i, N_{j-1}:N_j) =  \\
       A\left(M_{i-1}:M_i, K_0:K_1\right) \times B\left(K_0:K_1, N_{j-1}:N_j\right) + \dotsb + \\
       A\left(M_{i-1}:M_i, K_{k-1}:K_k\right) \times B\left(K_0:K_1, N_{j-1}:N_j\right)
\end{multline*}

Thus, block matrices can be multiplied in the same way as regular matrices,
by replacing each scalar multiply with a matrix multiply, and each scalar
addition with a matrix addition.

A special case of the above property allows us to partition matrices
$A'$ and $B'$ into panels, and matrix $C$ into blocks where the partitioning
of C dictates the partitioning of $A'$ and $B'$:

\begin{equation}
  \begin{aligned}
A' &= \left[
  \begin{array}{c}
  A_1 \\
  \vdots \\
  A_M
  \end{array}
\right]  \\
B' &= \left[
  \begin{array}{ccc}
  B_1 & \dotsb & B_N
  \end{array}
\right]  \\
C &= \left[
  \begin{array}{ccc}
    C_{11}  & \dotsb  & C_{1N}  \\
    \vdots & \ddots & \vdots \\
    C_{M1}  & \dotsb & C_{MN}  \\
  \end{array}
\right]
\end{aligned}
\end{equation}
Then each thread can compute a panel in the matrix multiply:
\[
  C_{ij} = \alpha \times A_i \times B_j + \beta \times C{ij}.
\]

\begin{figure}
  \begin{subfigure}{\columnwidth}
    \includegraphics[width=\linewidth]{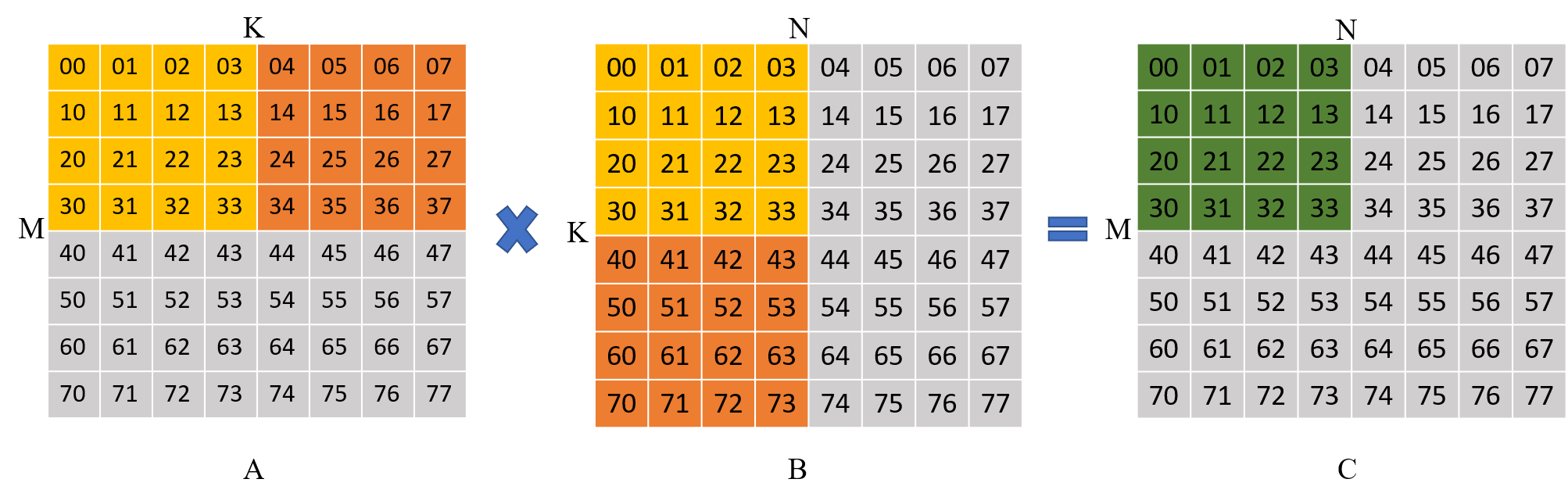}
    \Description{A matrix multiplication between two 8 by 8 matrices with the
    first 4 by 4 submatrix of the result highlighted, along with the
    corresponding panel of the first 4 rows of the left hand matrix and the
    first 4 columns of the right hand matrix which contribute to that output
    submatrix.}%
    \caption{global memory}\label{fig:matmul}
  \end{subfigure}
  \begin{subfigure}{\columnwidth}
    \centering
    \includegraphics[width=0.75\linewidth]{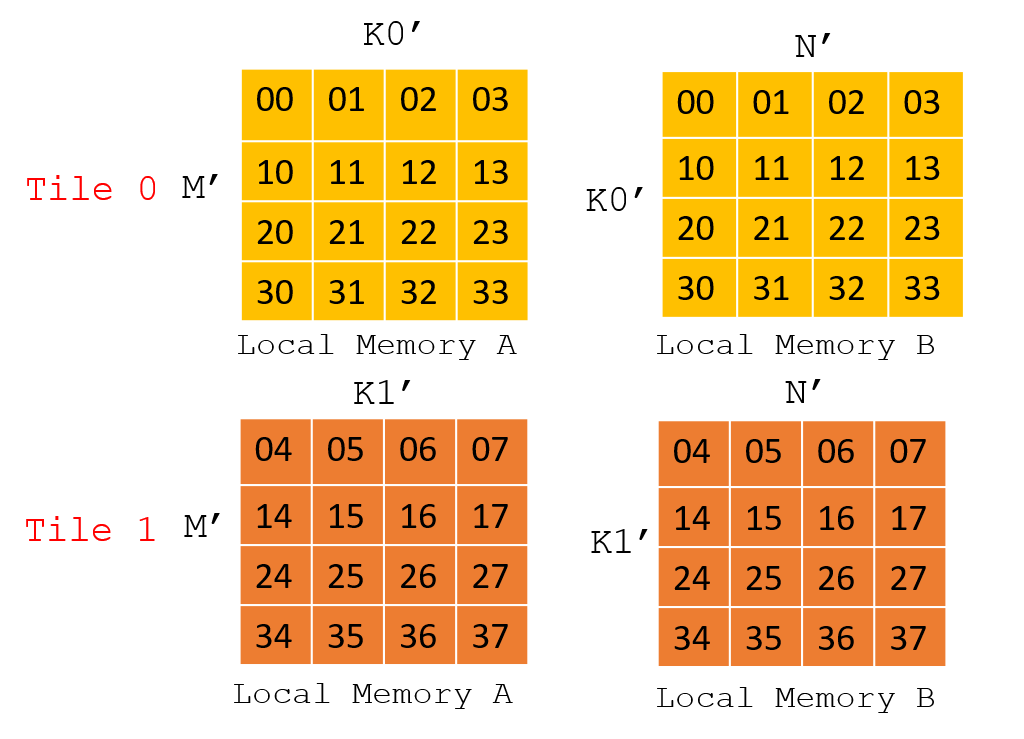}
    \Description{The highlighted panels from Figure~\ref{fig:matmul} of the two
    input matrices split into 4 by 4 blocks, containing the data required to
    compute a 4 by 4 block of the output matrix.}%
    \caption{local memory}\label{fig:matmullocal}
  \end{subfigure}
  \begin{subfigure}{\columnwidth}
    \includegraphics[width=\linewidth]{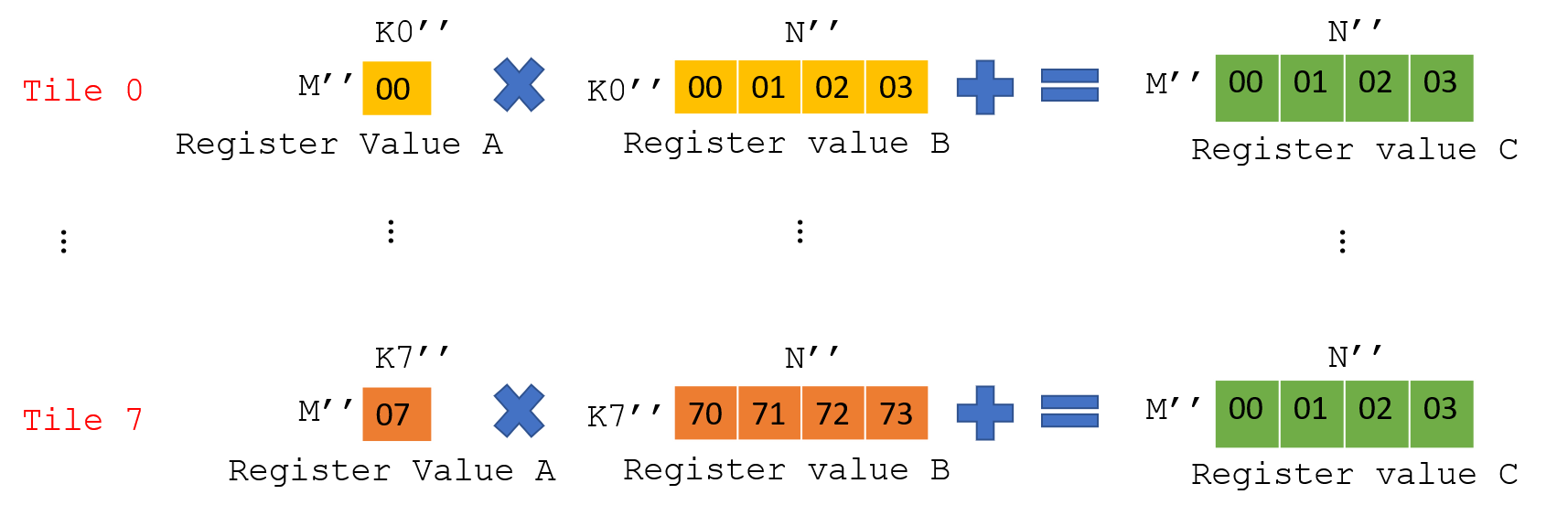}
    \Description{The blocks from Figure~\ref{fig:matmullocal} split into
    individual registers, showing the computation of 4 elements of the output.
    The first 4 output elements depend on the first row of the left hand matrix
    and on the first 4 columns of the right hand matrix. When individual values
    of the left hand matrix are loaded one by one there are 8 vector
    multiply-adds.}%
    \caption{private memory}\label{fig:matmulregister}
  \end{subfigure}
  \begin{subfigure}{\columnwidth}
    \includegraphics[width=\linewidth]{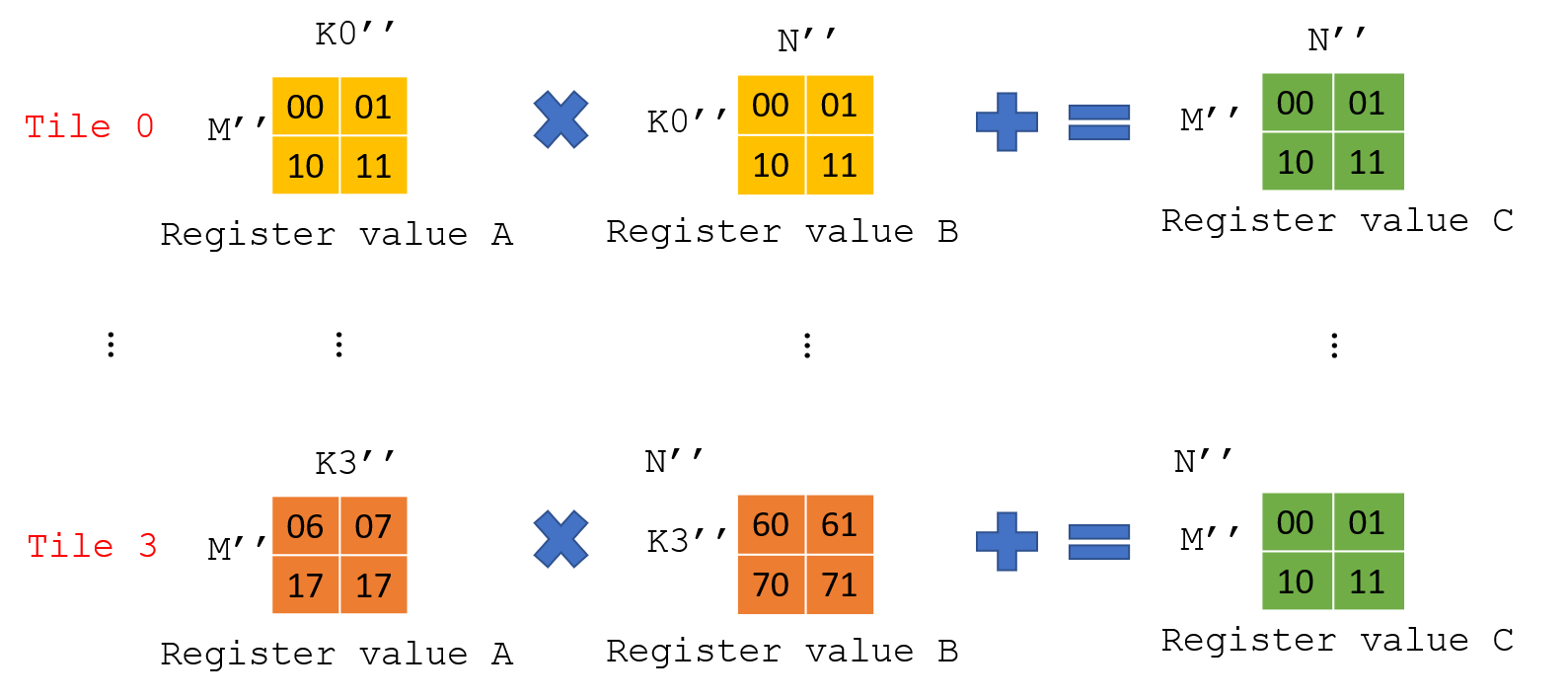}
    \Description{Similar to Figure~\ref{fig:matmulregister}, the block matrices
    are split into registers however here the left hand matrix is split into
    small 2 by 2 tiles in registers. This allows a 2 by 2 output tile to be
    computed over 4 tiles.}%
    \caption{private memory}\label{fig:matmulregister2}
  \end{subfigure}
  \caption{Tiled-based Matrix multiply}\label{fig:tiledmatmul}
\end{figure}

Each panel can then itself be broken down further into tiles and we recursively
build tiles of smaller matrices in different memory hierarchies. To illustrate
this tile based approach, suppose we are performing an $8 \times 8$ floating
point matrix multiply, for a device with 128 bytes of local memory and a maximum
of 12 float registers available without spilling.  Further suppose that 4
work-groups with 4 threads in each work-group are created.
Figure~\ref{fig:matmul} represents the $A'$ and $B'$ panels and $C$ blocks in
global memory.  The tile sizes are different in the memory hierarchies due to
memory size limitations.  The faster the memory is, the smaller the size of the
memory, and therefore the smaller the tile size. Thus multiple iterations are
required to load all the required data for the computation.

Since the total number of threads is 16, each thread computes 4 output elements.
At first all threads of a work-group load the first tile of the $A'$ and $B'$
panel (the yellow square in Figure~\ref{fig:matmul}) into local memory in a
coalesced way.  Given the size of the local memory, each input matrix must be
broken into two tiles (represented by yellow and orange color in
Figure~\ref{fig:matmul}) where one tile per matrix can be loaded into local
memory at once (Figure~\ref{fig:matmullocal}).  Similarly, each tile of the $A'$
and $B'$ panels in local memory is broken into 7 tiles of $A''$  and $B''$ to
compute the matrix multiply without register spill
(Figure~\ref{fig:matmulregister} and Figure~\ref{fig:matmulregister2}).

Moreover, there are different ways of loading tiles from local memory to
registers. Figure~\ref{fig:matmulregister} loads the local memory
tiles into 7 separate register tiles, which enables coalesced writes to
global memory. In Figure~\ref{fig:matmulregister2}, although the number of
tiles is reduced due to using more registers, each thread writes a block of
2 consecutive elements, thus the memory writes are not coalesced.

Choosing between the two above tile configurations at the private memory level
is platform specific. While in a SIMD platform coalesced read/write have a
significant effect on performance, on non-SIMD devices (e.g CPUs) block data
accesses and a reduced tile count provide better performance.

Providing a parametric GEMM enables a programmer to tune the operation for different
platforms by choosing different configurations without re-implementing the kernel.

\subsubsection{Data reuse in a GEMM operation}\label{sec:gemm_data_reuse}

Assuming a $m' \times n'$ submatrix $C_{ij}$ is already in registers, the
computation of a single block across $k'$ elements requires reading $m' k' + k'
n'$ data entries from global memory, and requires $2 m' n' k'$ floating point
operations.

Overall, $2 K m' n'$ floating point operations are needed to compute $C_{ij}$,
and $m' n' + m' K + n' K$ data elements are required to be loaded from global
memory.

The panels $A_i$ and $B_j$ are read in blocks of size $k'$, due to memory and
register limitations. Then each block of $A_i$ will be
multiplied with the appropriate block of $B_j$ and accumulated into $C_{ij}$.
In this way $C_{ij}$ is stored in registers during the entire operation and only
written to global memory at the end.

The amount of data reuse obtained through this tiled approach to
multiplying blocks is therefore given by:
\begin{equation}\label{eq:tilesize}
  \frac{2m' n' k'}{m'k' + k'n'} = \frac{2 m' n'}{m' + n'}.
\end{equation}

Since the goal of the blocking is to improve the amount of data reuse, the
block sizes should be chosen to maximize this property.

Equation~\ref{eq:tilesize} shows that increasing $k'$ does not increase the
amount of data reuse, but does increase the amount of register/local
memory/cache space required to store the matrices. For the tiles in private
memory $k' = 1$ seems to be the best choice.  Increasing $m'$ and $n'$ increases
data reusability, but it also increases the amount of local
memory/register/cache space that is required. It can easily be shown that, with
an upper limit on the available amount of memory, the best reuse is obtained if
$m' = n'$, i.e.\ if blocks $C_{ij}$ are chosen to be square.

When supported, using local memory will significantly improve
the performance when $A$ is transposed or $B$ is not transposed.
The other two possibilities may not require explicit copies from
global to local memory, as assigning items
to elements of the block in the correct order yields a perfectly
coalesced read of these matrices from memory. Such temporal locality
ensures that the data is already cached when a
different item of the same work group requests the same data entity.
However, moving the data explicitly to local memory can
improve performance when the hardware cache unit is a slower
than the local memory unit.

Software pre-fetching or double buffering is a well-known technique to
overcome the latency of off-chip memory accesses.
By doubling the size of local memory, it is possible to load
the next tile while computing the current tile. This can hide
the latency of the loads by computations.


\section{SYCL-DNN}\label{sec:sycldnn}

SYCL-DNN~\cite{sycldnn} is an open-source library which provides optimized
routines to accelerate neural network operations. It is built using the SYCL
programming model to enable cross platform support, and so works on any
accelerator supported by the user's chosen SYCL implementation.

Modern image recognition networks mainly involve many convolutions and matrix
multiplies, which make up the majority of the runtime of these networks. The
matrix multiplies can be supplied by a BLAS implementation like SYCL-BLAS\@. The
purpose of SYCL-DNN is to provide convolutions and other similar routines which
are required for these networks.

\subsection{Convolutions}\label{sec:conv}

Convolutions are typically the most compute intensive operation in these
networks, and so they are the most optimized in the library. To optimize this
operation, SYCL-DNN provide implementations of a number of algorithms which
compute a 2D convolution. These different algorithms have different performance
characteristics and memory requirements, and so behave differently for different
tensor sizes and on different devices.

In neural networks, a 2D convolution is an operation taking a batch of 3D input
tensors and a 4D filter tensor. Each output element is the sum of products of an
input value and a weight given by sliding a small window over the input tensor.

\newcommand{\inrows}{\ensuremath{H}}
\newcommand{\incols}{\ensuremath{W}}
\newcommand{\filterrows}{\ensuremath{R}}
\newcommand{\filtercols}{\ensuremath{S}}
\newcommand{\inchannels}{\ensuremath{C}}
\newcommand{\outchannels}{\ensuremath{K}}

More formally, let $\inrows$ be the number of rows in the input, $\incols$ the
number of columns, $\inchannels$ the number of input channels, $\outchannels$
the number of output channels and $\filterrows \times \filtercols$ the filter
size. Then the input tensor $I \in \mathbb{R}^{\inrows \times \incols \times
\inchannels}$ and the filter tensor $F \in \mathbb{R}^{\filterrows \times
\filtercols \times \inchannels \times \outchannels}$ would give rise to an
output tensor $O \in \mathbb{R}^{\inrows \times \incols \times \outchannels}$.

\begin{algorithm}
  \DontPrintSemicolon{}
  \KwData{input[\inrows][\incols][\inchannels]}
  \KwData{filter[\filterrows][\filtercols][\inchannels][\outchannels]}
  \KwData{output[\inrows][\incols][\outchannels]}
  \For{h = 1 \KwTo{} \inrows}{%
    \For{w = 1 \KwTo{} \incols}{%
      \For{k = 1 \KwTo{} \outchannels}{%
        out = 0\;
        \For{x = 1 \KwTo{} \filterrows}{%
          \For{y = 1 \KwTo{} \filtercols}{%
            \For{c = 1 \KwTo{} \inchannels}{%
              out += input[h + x][h + y][c] * filter[x][y][c][k]\;
            }
          }
        }
        output[h][w][k] = out\;
      }
    }
  }
  \caption{Naive algorithm to compute 2D convolution.}%
  \label{algo:conv2d}
\end{algorithm}

Using these layouts, the convolution can be computed naively as described in
Algorithm~\ref{algo:conv2d}. For example with a filter size of 3 $\times$ 3, an
output element at row $h$, column $w$ and channel $k$ would be computed as:
\[
  o_{h, w, k} = \sum_{c = 1}^{\inchannels} \sum_{x = 1}^{3} \sum_{y = 1}^{3}
    i^{h,w}_{x, y, c} \times f_{x, y, c, k}
\]
where $i^{h, w}$ is a $3 \times 3 \times \inchannels$ slice of the input tensor
centered at row $h$ and column $w$.

\subsubsection{Tiling 2D convolutions}\label{sec:tiledconv}

For convolutions with window sizes greater than $1$, there are overlaps in the
regions of the input tensor which affects two spatially adjacent output
elements. The naive method of computing a convolution using
Algorithm~\ref{algo:conv2d} on a GPU would use a single thread per output
element, which means every thread has to load all of the input slice that
it requires from memory.

\begin{figure*}[p]
  \includegraphics{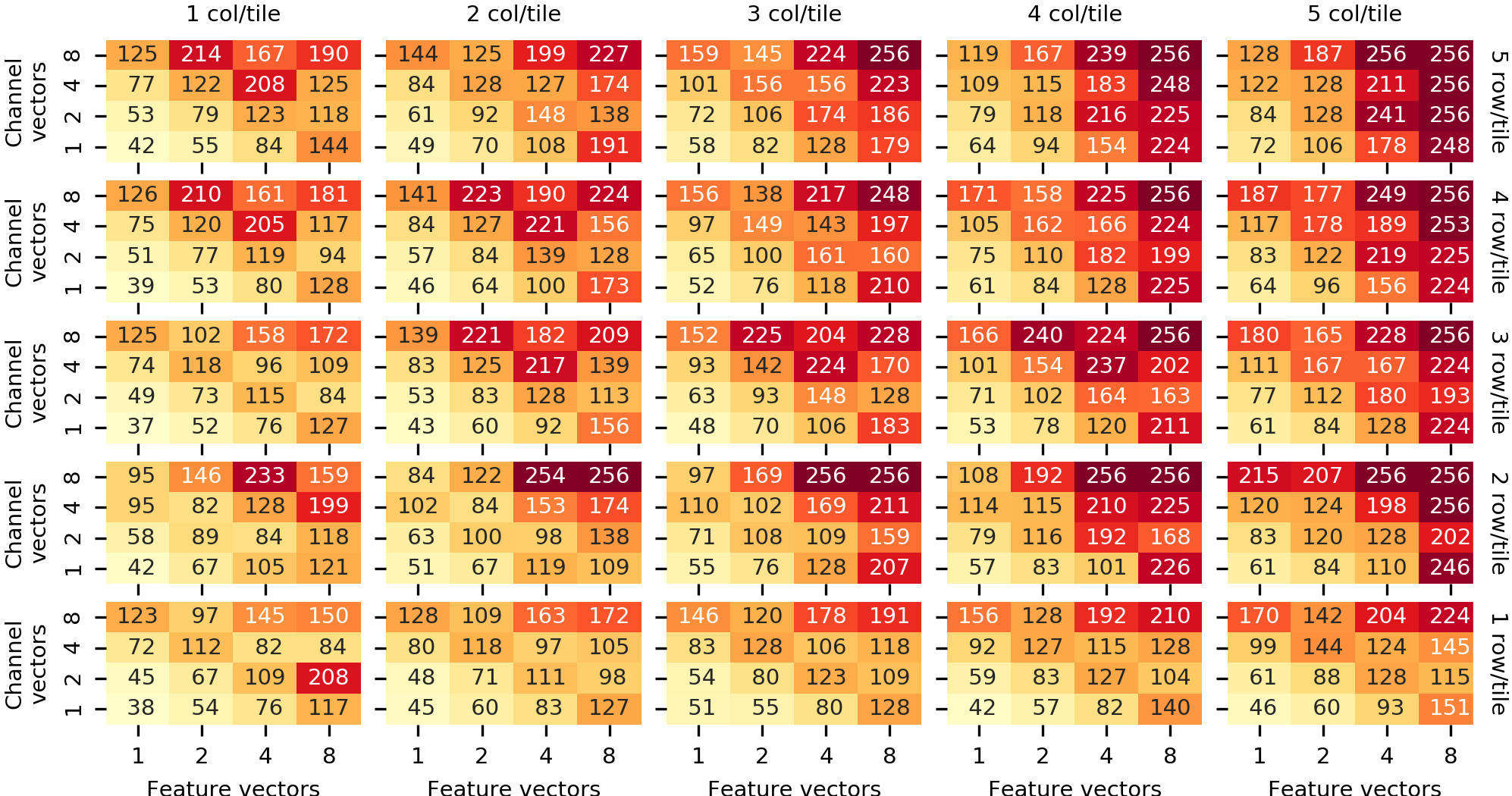}
  \Description{A grid of heatmaps where each heatmap corresponds to a given tile
  size, from 1 by 1 up to 5 by 5. Inside a given heatmap the entries are given
  by power of 2 vector sizes from 1 to 8. The number of registers increases as
  the tile increases and as the vector sizes increases from 38 registers in the
  naive case through to 256 in the largest tile sizes.}%
  \caption{Number of registers used for different tile sizes and vector sizes
  for SYCL-DNN's tiled 3$\times$3 convolution kernel.
  Each subplot gives the number of registers used for a given tile size as the
  vector sizes vary. Register usage data generated by AMD CodeXL~\cite{codexl}.}%
  \label{fig:reg_use}
\end{figure*}

\begin{figure*}[p]
  \includegraphics{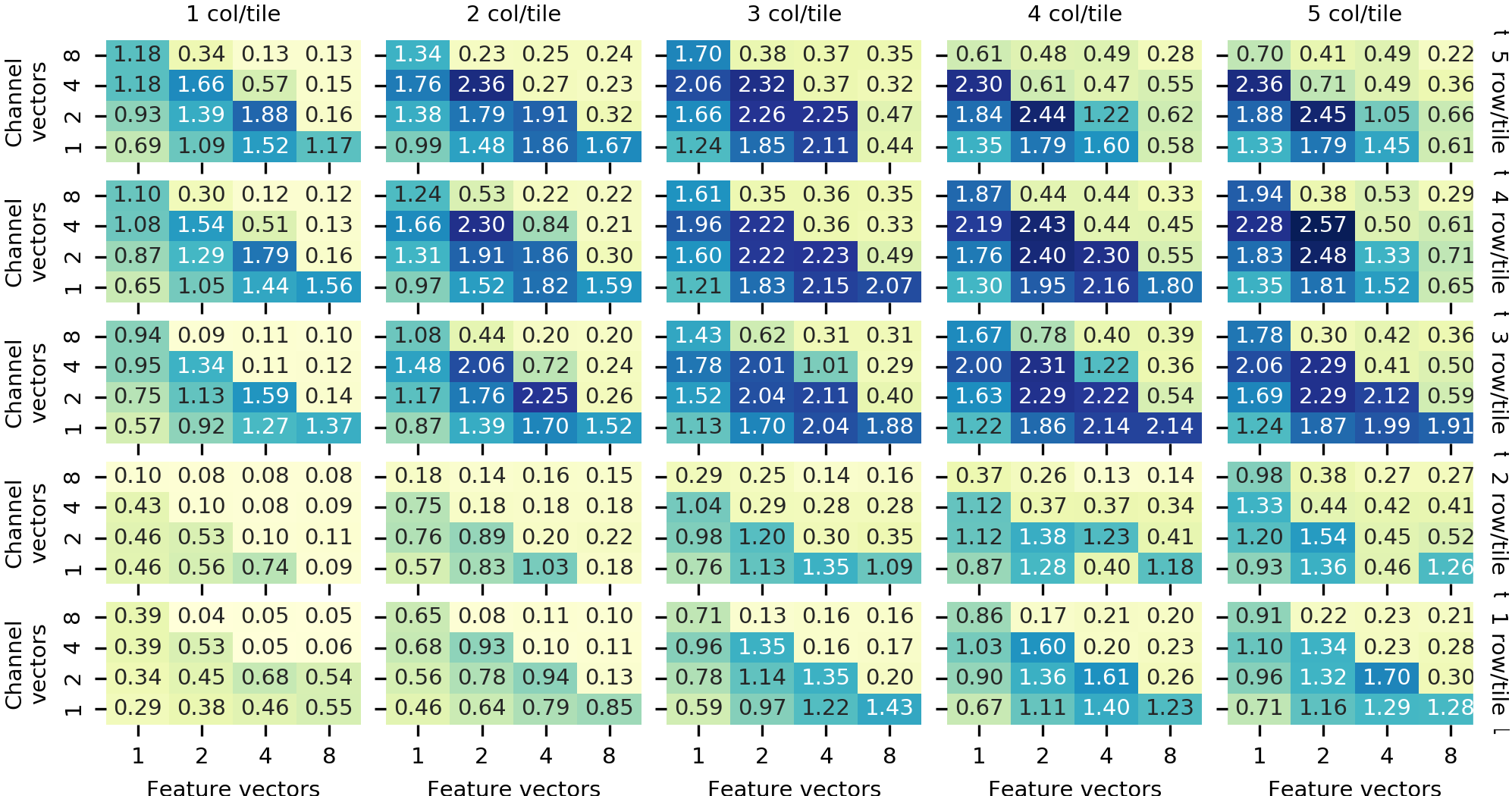}
  \Description{A comparison to Figure~\ref{fig:reg_use} where instead of
  register usage, the achieved performance is shown. The number of flops
  increases as tile sizes increase and as vector sizes increase, however in each
  heatmap there is a point where increasing vector sizes causes the performance
  to drop significantly. This often happens when using 8 element vectors.}%
  \caption{Number of teraflops achieved for different tile sizes and vector
  sizes for a given 3$\times$3 convolution.
  Each subplot gives the performance for a given tile size as the vector sizes
  vary. Performance data obtained using an AMD R9 Nano GPU, with amdgpu-pro
  divers.}%
  \label{fig:tiled_perf}
\end{figure*}

Adjacent elements in the convolution output require overlapping slices of the
input tensor, so using a single thread to compute a number of these elements
will reduce the number of times a single input element is loaded from memory. We
refer to this as a tiled algorithm, as each thread will compute a tile of the
output. As the tile size increases the number of times each input element is
reused also increases and so the total number of bytes read from memory in the
whole operation decreases. Memory bandwidth is typically the largest bottleneck
in these GPU computations, as discussed in Section~\ref{sec:perfmetrics}, and so
reducing the number of loads typically improves performance accordingly.
Similarly the use of vector loads and stores, as well as vector math operations,
can improve performance on some devices.

The downside of using large tiles and vectors in GPU kernels is that each
significantly increases the number of registers required to hold the data
required for a single thread. As an example, Figure~\ref{fig:reg_use} shows the
number of registers required for varying tile and vector sizes as provided by
AMD's CodeXL tools~\cite{codexl}.

As GPUs have a limited number of registers available in their register files,
the number of registers required by a single thread determines how many threads
can be concurrently executed.  Overhead and memory fetches are largely hidden in
a GPU by quickly switching threads when one becomes stalled. If each thread
requires more registers then the number of concurrent threads decreases and so
the chance that one is ready to execute as soon as another is stalled is also
decreased.

The best performance then does not necessarily come when using the largest tile
sizes to get the most data reuse and large vector sizes to get more efficient
loads and instruction level parallelism, but also the best performance is not
achieved when using the fewest registers and so allowing the GPU to run the most
threads concurrently.

Optimal performance is typically achieved by balancing these three factors, as
shown in Figure~\ref{fig:tiled_perf} where the peak performance on the AMD R9
Nano is achieved using a $4 \times 5$ tile, $4$ element vectors for input
channels and $2$ element vectors for output channels (at $2.57$ teraflops),
which gives a $10\times$ speed up over the naive implementation (at $0.29$
teraflops). When a tile size which uses more registers than are available on the
device---and so any values in the registers are instead stored in memory---the
performance is worse still, going down to as little as $50$ gigaflops.

On different devices, the optimal tile sizes and vector widths to get best
performance will differ depending on the device's load-store units, the number
of available registers and whether vector math units are available.

\begin{table}
  \caption{SYCL-BLAS configuration for local and register tile sizes.
  `Registers' shows the number of registers in the tile, `Work group' shows
  the work-group size used and `Local mem' refers to the size of local memory
  used.}%
  \label{tbl:gemm-config}
  \begin{tabular}{lccr}
    \toprule
    \textbf{Configuration} & \textbf{Registers} & \textbf{Work group}&\textbf{Local mem} \\
    \midrule
    \texttt{4$\times$4\_8$\times$8\_loc}      &  16  &  64 &  8 KiB  \\
    \texttt{4$\times$4\_16$\times$16\_loc}    &  16  & 256 & 16 KiB  \\
    \texttt{8$\times$4\_8$\times$16\_loc}     &  32  & 128 & 16 kiB  \\
    \texttt{8$\times$2\_4$\times$16\_loc}     &  16  &  64 &  8 KiB  \\
    \texttt{8$\times$4\_8$\times$16\_noloc}   &  32  & 128 & N/A     \\
    \texttt{8$\times$4\_4$\times$8\_noloc}    &  32  &  32 & N/A     \\
    \texttt{4$\times$4\_8$\times$8\_noloc}    &  16  &  64 & N/A     \\
    \bottomrule
  \end{tabular}
\end{table}

\subsubsection{Winograd technique for fast convolutions}\label{sec:winograd}

Lavin and Gray suggested a technique to accelerate convolutions
in~\cite{winograd} by using a transform studied by
Winograd~\cite{winograd_complexity} as well as Cook and Toom. In many DNN
libraries this is referred to as the Winograd technique.

The technique involves transforming the input and filter tensors into a number
of small matrices, then the convolution is computed using a batched matrix
multiply between these matrices, followed by another transform to yield the
output values. In this way the total number of floating point operations
required to compute the convolution can be decreased to as little as $30\%$ of
that required by a naive computation.

For a convolution with a $R \times S$ filter size, the input transform converts
an $M \times N$ tile of the input tensor to a $(M + R - 1) \times (N + S - 1)$
tile, which is then scattered across $(M + R - 1) (N + S - 1)$ matrices with one
tile element equal to an corresponding element in one of the matrices.

Increasing the tile sizes used in the transforms increases the amount of data
reuse, so overall fewer memory fetches are required as discussed in
Section~\ref{sec:tiledconv}. However larger tile sizes also require more
registers as above, which can lead to fewer concurrent threads and possibly
worse performance.

Another effect of using larger tiles sizes is that the number of intermediate
matrices increases, but the size of each individual matrix decreases. The
majority of the computation is the batched matrix multiply, and for smaller
matrices it can be harder to fully utilize a GPU even with specialized batched
kernels.

Given these considerations, there is again the problem of balancing the tile
size with register usage, optimal matrix multiply sizes and vector widths in
order to get the best performance. As with the tiled convolution computations,
this will differ depending on the device as well as the optimizations available
in the matrix multiplies. The performance portability provided by the SYCL-BLAS
matrix multiplies discussed in Section~\ref{sec:syclblas} significantly affects
the achievable performance when using this technique to compute convolutions.

\section{Performance}

The performance of our SYCL kernels is measured against hand-tuned, vendor
provided libraries.  The primary target devices of our libraries have been
embedded devices, so the main comparison has been made to ARM's
ComputeLibrary~\cite[v18.11]{acl}. This provides accelerated routines through both NEON
instructions for the CPU and OpenCL kernels for GPUs.

We also provide comparisons for Intel hardware, utilizing both the CPU and GPU,
with comparisons to Intel MKL-DNN~\cite[v0.18.1]{mkldnn} and to
clBLAST~\cite{nugteren2018clblast}.

\subsection{Benchmark device setup}

The performance benchmarks are provided on a range of hardware, to show the
portability of our SYCL libraries.

\subsubsection{HiSilicon Kirin 960}

The HiSilicon HiKey 960 is a development board featuring the HiSilicon Kirin 960
system on a chip (SoC) designed for use in mobile phones. This comprises an ARM
big.LITTLE CPU with 4 ARM A73 cores and 4 ARM A53 cores, along with an
ARM Mali G-71 GPU\@.

When benchmarking using the GPU, the big (A73) CPUs are disabled to help keep
device temperatures low. For benchmarking on the CPU, the big A73 cores are
enabled.

\subsubsection{Intel i7-6700K}

The i7-6700K is a Skylake desktop processor which contains 4 CPU cores with
a base frequency of 4GHz, boosting to 4.2GHz, and an Intel HD Graphics 530
integrated GPU with a base frequency of 350 MHz, boosting to 1.15GHz.

For the neural network benchmarks in Section~\ref{sec:nn_bench}, we benchmarked
SYCL-DNN using both the CPU and GPU in the i7-6700K processor, as Intel
provide an OpenCL implementation for Xeon and Core CPUs and an open source
Graphics Compute Runtime for their integrated graphics processors.  When
benchmarking using the CPU the scaling governor was set to `performance'.

\subsubsection{Intel i7-9700K}

The i7-9700K is a Skylake desktop processor which contains 8 CPU cores with
a base frequency of 3.6GHz, boosting to 4.9GHz, and an Intel UHD Graphics 630
integrated GPU with a base frequency of 350 MHz, boosting to 1.2GHz.

For the BLAS benchmarks in Section~\ref{sec:blas_gemm_bench}, we benchmarked
only the GPU in the i7-9700K processor.

\subsection{SYCL-BLAS GEMM benchmarks}\label{sec:blas_gemm_bench}

\begin{figure}
  \begin{subfigure}{\columnwidth}
    \includegraphics{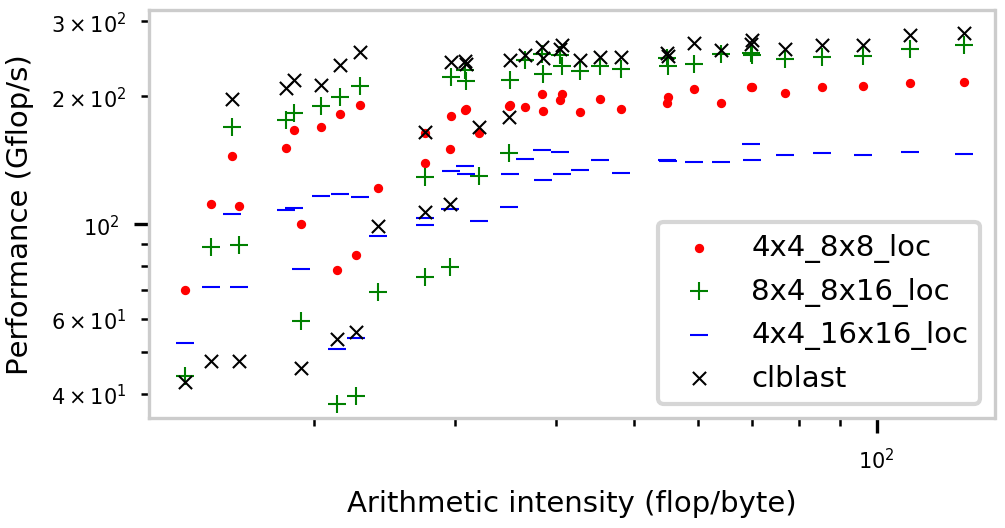}
    \Description{A scatter plot showing a roofline comparison of the arithmetic
    intensity in operations per byte with the total performance in operations
    per second. Each matrix multiply is computed with different configurations
    and compared to clBLAST\@. The clBLAST figures perform best, though only
    around 20 gigaflops ahead of one of the SYCL-BLAS configurations. The other
    configurations perform worse.}%
    \caption{Overall performance comparison of SYCL-BLAS for a number of tile
    configurations against hand-tuned clBLAST.}%
    \label{fig:intel-all}
  \end{subfigure}
  \begin{subfigure}{\columnwidth}
    \includegraphics{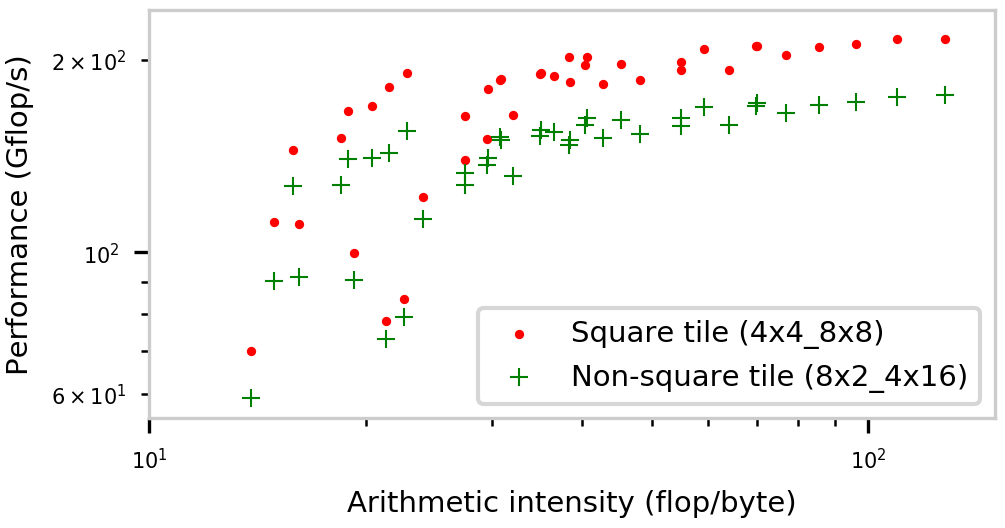}
    \Description{A roofline comparison between two SYCL-BLAS configurations one
    with a square 4 by 4 register tile and the other with an 8 by 2 tile. The
    square tile performs best in all cases.}%
    \caption{Performance comparison of using a square register tile vs a
    non-square register tile, when the total number of registers is constant.}%
    \label{fing:intel-square}
  \end{subfigure}
  \begin{subfigure}{\columnwidth}
    \includegraphics{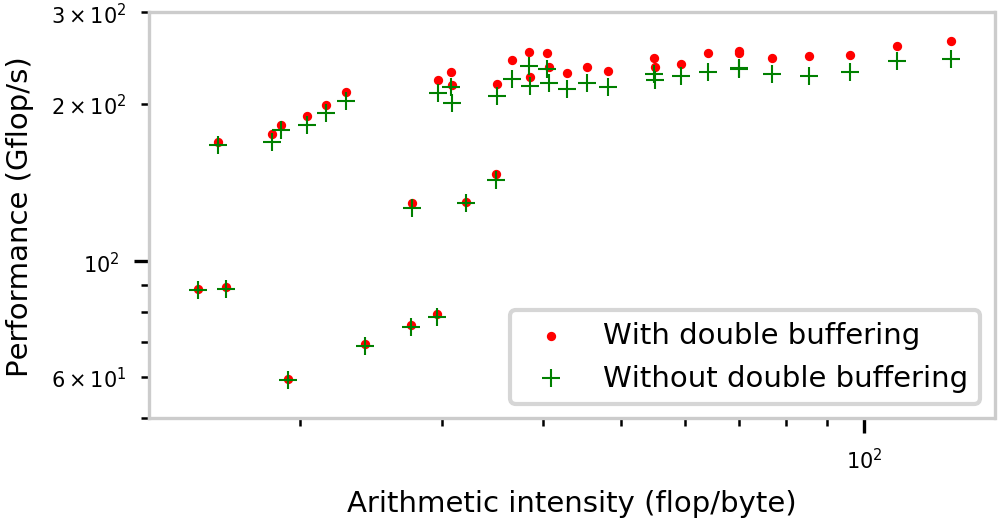}
    \Description{A roofline comparison between the same SYCL-BLAS configuration
    with and without double buffering. Using double buffering gives better
    performance in all cases shown.}%
    \caption{Performance comparison of using double buffering. Figures shown
    with configuration \texttt{8$\times$4\_8$\times$16\_loc}.}%
    \label{fig:intel-double-buffer}
  \end{subfigure}
  \caption{Comparison of different performance metrics of SYCL-BLAS on an Intel
  UHD Graphics 630 GPU.}%
  \label{fig:intel-gemm}%
\end{figure}

A comparison similar to the roofline model~\cite{williams2009roofline}
is carried out to plot performance of a kernel (Gflop/s) as a function of its operational intensity,
i.e the number of floating point operations per byte of data read or written (flop/byte).

The SYCL-BLAS GEMM implementation has been compared to clBLAST's hand-tuned GEMM
OpenCL kernels~\cite{nugteren2018clblast} on an Intel UHD Graphics 630 GPU and
to ARM Compute Library's GEMM kernels on an ARM Mali G-71 GPU\@.  For each
platform, different configurations of SYCL-BLAS have been used, with various
register tile sizes, work-group sizes, and local memory sizes, as shown in
Table~\ref{tbl:gemm-config}. Each configuration contains
\texttt{h$\times$w\_r$\times$c} where
\texttt{h$\times$w} represents a tile of $h$ by $w$ registers per thread to compute the
block matrix $C_{ij}$,  and \texttt{r$\times$c} represents a work-group consisting of
$r \times c$ threads.

The number of data elements stored in local memory for the computation of a
matrix multiply between matrices of size $M \times K$ and $K \times N$, for a
configuration \texttt{hxw\_rxc} is given by:
\[ h \times r \times X  +  X \times w \times c \]
where $X$ is the number of data elements that fit within a cache line.  The
local memory size is doubled when double buffering is used. In
Figures~\ref{fig:intel-gemm} and~\ref{fig:arm-gemm} each point represents the
performance of a single matrix multiply between matrices of size $M\times K$ and
$K\times N$, where $M \in [64, 1204]$, $N \in [64,1024]$, $K \in [64, 1024]$ and
$M$, $N$, and $K$ are all powers of 2.

\subsubsection{Performance on Intel i7-9700K}

Figure~\ref{fig:intel-all} shows a roofline model comparing different
configurations of SYCL-BLAS GEMM implementation against the GEMM implementation
from clBLAST~\cite{nugteren2018clblast}.  Comparing these results shows that the
\texttt{8$\times$4\_8$\times$16\_loc} configuration for SYCL-BLAS achieves close
to the performance of the hand-tuned clBLAST OpenCL kernels on the Intel GPU\@.
In addition, the results show that increasing the number of registers from
$4\times 4$ to $8\times 4$ per thread significantly improves performance, as
data reuse is increased.

Figure~\ref{fing:intel-square} shows a roofline model comparing two different
SYCL-BLAS GEMM configurations which use 16 registers, but in different tile
layouts.  The result shows that the square register tile
\texttt{4$\times$4\_8$\times$8} yields better performance than the non-square
register tile \texttt{8$\times$2\_4$\times$16}, as discussed in
Section~\ref{sec:gemm_data_reuse}.  Similarly,
Figure~\ref{fig:intel-double-buffer} shows the performance improvement from
using double buffering to hide memory latency.

\subsubsection{Performance on ARM HiKey 960}

\begin{figure}
  \begin{subfigure}{\columnwidth}
    \includegraphics{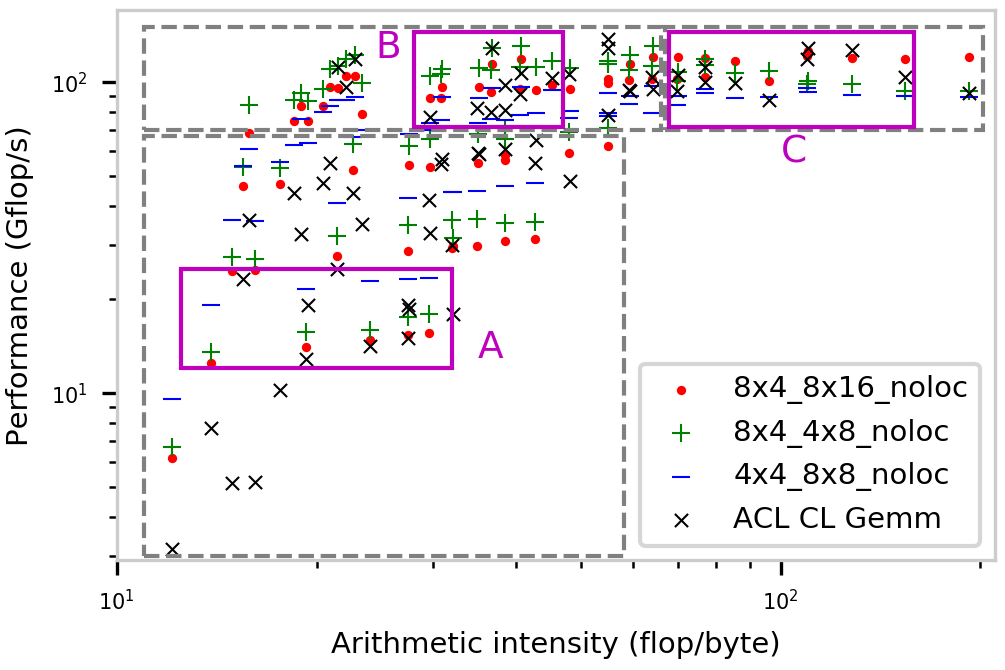}
    \Description{A large scatter plot showing a roofline comparison of 3
    SYCL-BLAS configurations with ARM Compute Library. There are three
    highlighted zones, the one marked A is in the lower left corner, one marked
    B is in the upper center and one marked C is in the upper right corner.}%
    \caption{Overall comparison of SYCL-BLAS against the hand-tuned OpenCL
    kernels in ARM's Compute Library across a variety of matrix sizes. The
    highlighted areas are shown in more detail below.}%
    \label{fig:arm-all}
  \end{subfigure}
  \begin{subfigure}{\columnwidth}
    \includegraphics{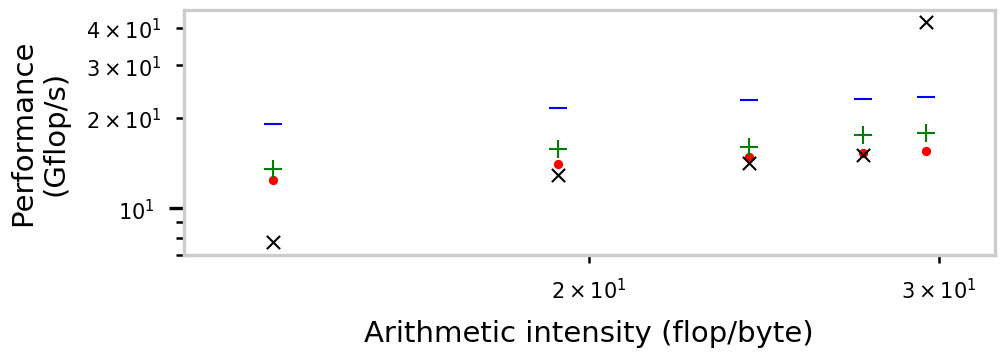}
    \Description{A more detailed view of zone A from Figure~\ref{fig:arm-all}
    showing that the SYCL-BLAS configuration with 4 by 4 register tiles performs
    the best, achieving around 20 gigaflops whereas ARM Compute Library achieves
    between 8 and 16.}%
    \caption{Block A, where \texttt{4$\times$4\_8$\times$8} achieves the best performance.}%
    \label{fig:arm-group1}
  \end{subfigure}
  \begin{subfigure}{\columnwidth}
    \includegraphics{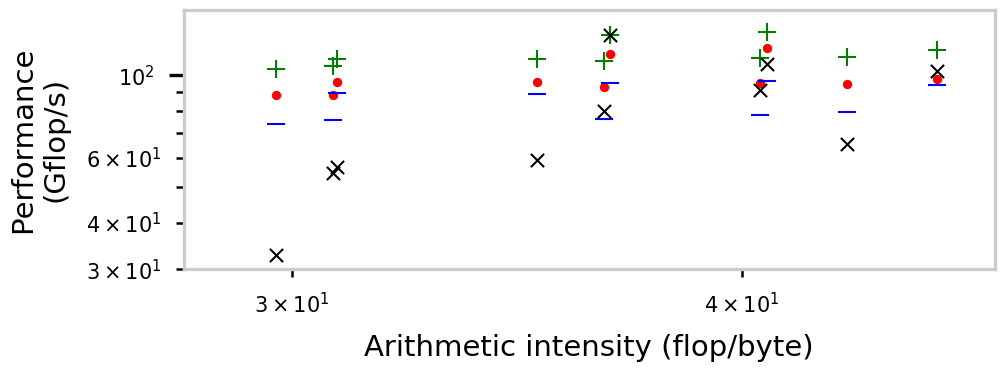}
    \Description{A more detailed view of zone B from Figure~\ref{fig:arm-all}
    showing that the SYCL-BLAS configuration with 8 by 4 register tiles performs
    the best, achieving between 100 and 140 gigaflops.}%
    \caption{Block B where \texttt{8$\times$4\_4$\times$8} achieves the best performance.}%
    \label{fig:arm-group2}
  \end{subfigure}
  \begin{subfigure}{\columnwidth}
    \includegraphics{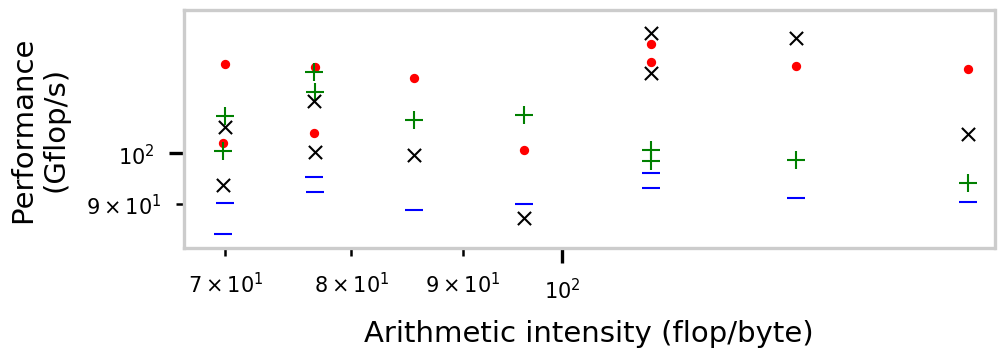}
    \Description{A more detailed view of zone C from Figure~\ref{fig:arm-all}
    showing that the SYCL-BLAS configuration with 8 by 4 register tiles but a
    larger work-group size performs the best in most cases, achieving between
    100 and 140 gigaflops. In these cases with larger matrices, ARM Compute
    Library is more competitive and in some of the cases outperforms SYCL-BLAS.}%
    \caption{Block C where \texttt{8$\times$4\_8$\times$16} achieves the best performance.}%
    \label{fig:arm-group3}
  \end{subfigure}
  \caption{Roofline performance comparison of SYCL-BLAS on an ARM Mali G-71 GPU,
  compared to ARM's Compute Library.}%
  \label{fig:arm-gemm}%
\end{figure}

Figure~\ref{fig:arm-all} shows a roofline model comparing the number of
gigaflops achieved by different configurations of SYCL-BLAS GEMM with that
achieved by hand-tuned OpenCL kernels in ARM Compute Library.  The figure is
divided in 3 dotted regions where different GEMM configurations give better
performance. These regions are shown in more detail in
Figures~\ref{fig:arm-group1} to~\ref{fig:arm-group3}.

In the region marked A, shown in Figure~\ref{fig:arm-group1}, the matrices have
a small size and are typically square, so the matrix multiply has a low
computational intensity and so comparatively few floating point operations. The
\texttt{4$\times$4\_8$\times$8} configuration shows the best performance for
SYCl-BLAS, and is competitive with ARM Compute Library.  However, in the region
marked B, Figure~\ref{fig:arm-group2}, the \texttt{8$\times$4\_4$\times$8}
configuration is better.  Here the matrices are small to medium sized and
typically rectangular, so the operation has a higher computational intensity but
still comparatively few operations, so launching fewer threads with a higher
workload works well.

Figure~\ref{fig:arm-group3} shows the region marked as C in more detail, where
the matrices are large and so the operation has high computational intensity and
many floating point operations.  In this region the
\texttt{8$\times$4\_16$\times$8} configuration is best, as the larger matrices
can be split up much better across a larger number of threads.

\subsection{Neural network benchmarks}\label{sec:nn_bench}

\begin{figure*}
  \includegraphics{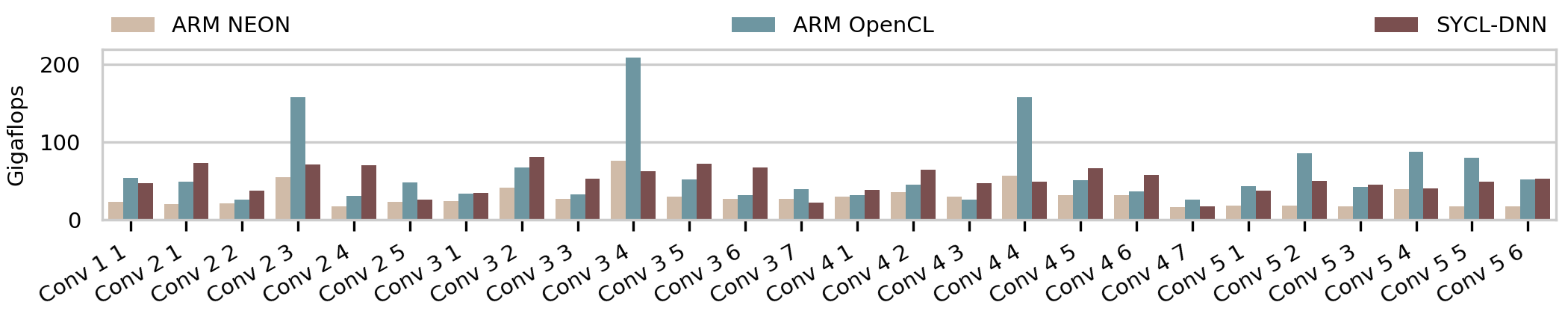}
  \Description{A bar chart with 26 groups of convolutions. In most cases
  SYCL-DNN performs better than ARM Compute Library by around 20 gigaflops,
  achieving between 40 and 80 gigaflops. The NEON implementation performs worst
  in almost all cases. There are three convolutions, corresponding to those with
  window size 3, where ARM Compute Library performs significantly better with
  between 150 and 200 gigaflops.}%
  \caption{Gigaflop comparison for ResNet layers on ARM Hikey 960 with SYCL-DNN
  and ARM Compute Library on both CPU using NEON and GPU using OpenCL\@. Benchmark
  run with a batch size of 1.}%
  \label{fig:arm_resnet}%
\end{figure*}

\begin{figure*}
  \includegraphics{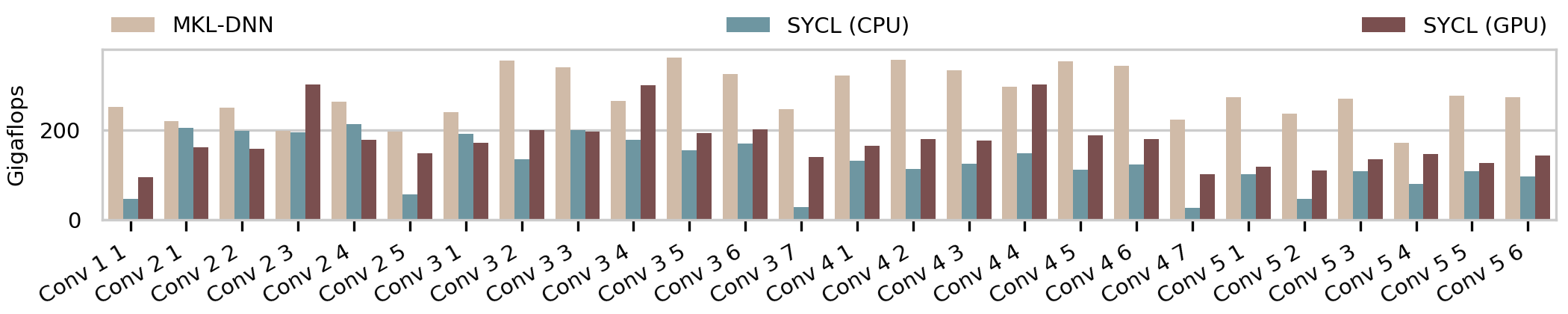}
  \Description{A bar chart with 26 groups of convolutions. In most cases MKL-DNN
  performs better than SYCL-DNN, achieving between 200 to 400 gigaflops, while
  SYCL-DNN can only achieve 150 to 300 gigaflops.}%
  \caption{Gigaflop comparison for ResNet layers on Intel i7-6700K with SYCL-DNN
   on both CPU and GPU compared to MKL-DNN on the CPU\@. Benchmark run with a
   batch size of 4.}%
  \label{fig:intel_resnet}%
\end{figure*}

\begin{table}
  \caption{Convolution sizes and parameters for each of the distinct layers in
  the VGG neural network model.
  The W column refers to the window size, S to the stride.}%
  \label{tab:vgg_params}%
  \begin{filecontents*}{data/vgg_layers.csv}
Layer,Window,Stride,Input,Output
Conv 1 1, 3, 1, 224 x 224 x   3, 224 x 224 x  64
Conv 1 2, 3, 1, 224 x 224 x  64, 224 x 224 x  64
Conv 2 1, 3, 1, 112 x 112 x  64, 112 x 112 x 128
Conv 2 2, 3, 1, 112 x 112 x 128, 112 x 112 x 128
Conv 3 1, 3, 1,  56 x  56 x 128,  56 x  56 x 256
Conv 3 2, 3, 1,  56 x  56 x 256,  56 x  56 x 256
Conv 4 1, 3, 1,  28 x  28 x 256,  28 x  28 x 512
Conv 4 2, 3, 1,  28 x  28 x 512,  28 x  28 x 512
Conv 5 1, 3, 1,  14 x  14 x 512,  14 x  14 x 512
\end{filecontents*}
\pgfplotstabletypeset[
    col sep=comma,
    columns={Layer,Window,Stride,Input,Output},
    string replace*={_}{\_},
    every head row/.style={before row=\toprule,after row=\midrule},
    columns/Layer/.style={%
      column type={l},
      string type
    },
    columns/Window/.style={%
      column name={W},
    },
    columns/Stride/.style={%
      column name={S},
    },
    columns/Input/.style={%
      string replace*={x}{$\times$},
      string type
    },
    columns/Output/.style={%
      string replace*={x}{$\times$},
      string type
    },
    assign column name/.style={/pgfplots/table/column name={\textbf{#1}}},
    every last row/.style={after row=\bottomrule},
  ]{data/vgg_layers.csv}%
\end{table}

\begin{table}
  \caption{Convolution sizes and parameters for each of the distinct layers in
  the ResNet neural network model.
  The W column refers to the window size, S to the stride.}%
  \label{tab:resnet_params}%
  \begin{filecontents*}{data/resnet_layers.csv}
Layer,Window,Stride,Input,Output
Conv 1 1, 7, 2, 230 x 230 x    3,  112 x 112 x  64
Conv 2 1, 1, 1,  56 x  56 x   64,   56 x  56 x 256
Conv 2 2, 1, 1,  56 x  56 x   64,   56 x  56 x  64
Conv 2 3, 3, 1,  56 x  56 x   64,   56 x  56 x  64
Conv 2 4, 1, 1,  56 x  56 x  256,   56 x  56 x  64
Conv 2 5, 3, 2,  56 x  56 x   64,   28 x  28 x  64
Conv 3 1, 1, 1,  28 x  28 x   64,   28 x  28 x 256
Conv 3 2, 1, 1,  28 x  28 x  256,   28 x  28 x 512
Conv 3 3, 1, 1,  28 x  28 x  256,   28 x  28 x 128
Conv 3 4, 3, 1,  28 x  28 x  128,   28 x  28 x 128
Conv 3 5, 1, 1,  28 x  28 x  128,   28 x  28 x 512
Conv 3 6, 1, 1,  28 x  28 x  512,   28 x  28 x 128
Conv 3 7, 3, 2,  28 x  28 x  128,   14 x  14 x 128
Conv 4 1, 1, 1,  14 x  14 x  128,   14 x  14 x 512
Conv 4 2, 1, 1,  14 x  14 x  512,   14 x  14 x 1024
Conv 4 3, 1, 1,  14 x  14 x  512,   14 x  14 x 256
Conv 4 4, 3, 1,  14 x  14 x  256,   14 x  14 x 256
Conv 4 5, 1, 1,  14 x  14 x  256,   14 x  14 x 1024
Conv 4 6, 1, 1,  14 x  14 x 1024,   14 x  14 x 256
Conv 4 7, 3, 2,  14 x  14 x  256,    7 x   7 x 256
Conv 5 1, 1, 1,   7 x   7 x  256,    7 x   7 x 1024
Conv 5 2, 1, 1,   7 x   7 x 1024,    7 x   7 x 2048
Conv 5 3, 1, 1,   7 x   7 x 1024,    7 x   7 x 512
Conv 5 4, 3, 1,   7 x   7 x  512,    7 x   7 x 512
Conv 5 5, 1, 1,   7 x   7 x  512,    7 x   7 x 2048
Conv 5 6, 1, 1,   7 x   7 x 2048,    7 x   7 x 512
\end{filecontents*}
\pgfplotstabletypeset[
    col sep=comma,
    columns={Layer,Window,Stride,Input,Output},
    string replace*={_}{\_},
    every head row/.style={before row=\toprule,after row=\midrule},
    columns/Layer/.style={%
      column type={l},
      string type
    },
    columns/Window/.style={%
      column name={W},
    },
    columns/Stride/.style={%
      column name={S},
    },
    columns/Input/.style={%
      string replace*={x}{$\times$},
      string type
    },
    columns/Output/.style={%
      string replace*={x}{$\times$},
      string type
    },
    assign column name/.style={/pgfplots/table/column name={\textbf{#1}}},
    every last row/.style={after row=\bottomrule},
  ]{data/resnet_layers.csv}%
\end{table}

\begin{figure}[tbp]
  \includegraphics{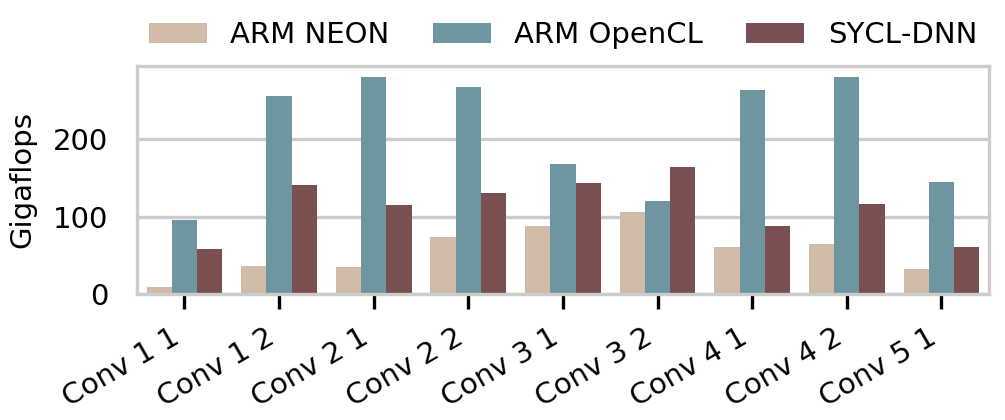}%
  \Description{A bar chart showing the 9 convolutions in VGG\@. In all but one
  case ARM Compute Library performs best. SYCL-DNN achieves between 90 and 150
  gigaflops, which ARM Compute Library varies more achieving between 100 and
  280. The NEON implementation performs the worst reaching a maximum of around
  100 gigaflops.}%
  \caption{Gigaflop comparison for VGG layers on ARM Hikey 960 with SYCL-DNN
  and ARM Compute Library on both CPU using NEON and GPU using OpenCL\@. Benchmark
  run with a batch size of 1.}%
  \label{fig:arm_vgg}%
\end{figure}

\begin{figure}[tbp]
  \includegraphics{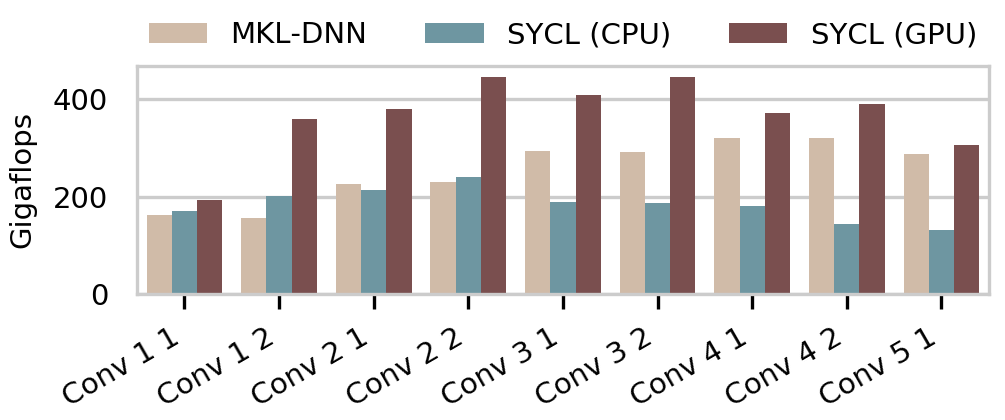}%
  \Description{A bar chart showing the 9 convolutions in VGG\@. SYCL-DNN
  performs the best using the GPU, achieving 200 gigaflops in the first
  convolution and between 300 and 420 in the others. MKL-DNN achieves between
  180 and 300 gigaflops.}%
  \caption{Gigaflop comparison for VGG layers on Intel i7-6700K with SYCL-DNN
   on CPU and GPU compared to MKL-DNN on the CPU\@. Benchmark run with a
   batch size of 4.}%
  \label{fig:intel_vgg}%
\end{figure}

To determine the performance of SYCL-DNN compared to other neural network
acceleration libraries we extracted the convolution layers from well known
standard image recognition networks. The networks chosen are VGG-16~\cite{vgg}
and ResNet-50~\cite{resnet}, introduced in 2014 and 2015 respectively, which are
fairly large convolutional networks which were state-of-the-art when introduced,
and are still frequently used as they provide good results and are widely
available.

The VGG model uses 16 layers of 3$\times$3 convolutions interspersed with max
pooling layers, while the ResNet model is made up of a number of blocks each
consisting of a 1$\times$1 convolution, a 3$\times$3 convolution then a
1$\times$1 convolution. Exact sizes for these layers as used in the benchmarks
are shown in Table~\ref{tab:vgg_params} and~\ref{tab:resnet_params}.

The performance results for the HiKey 960 SoC are shown in
Figures~\ref{fig:arm_vgg} and~\ref{fig:arm_resnet} for the VGG and ResNet
benchmarks respectively. SYCL-DNN is competitive with ARM's compute library and
typically out performs both the OpenCL and Neon implementations in the ResNet
benchmarks. The VGG model makes use of more 3$\times$3 convolutions, which have
very optimized OpenCL implementations in the ARM library and in most cases
outperform SYCL-DNN\@. These 3$\times$3 convolutions are also the three cases that
stand out in the ResNet benchmark where the ARM OpenCL performance is
significantly higher.

Figures~\ref{fig:intel_vgg} and~\ref{fig:intel_resnet} show the performance of
the VGG and ResNet benchmarks on the Intel i7-6700K platform, comparing the
performance of SYCL-DNN on both the CPU and the GPU against MKL-DNN\@. For the
convolutions in the ResNet model MKL-DNN is consistently faster than
SYCL-DNN, achieving up to 366 gigaflops while SYCL-DNN achieves a maximum of
244. However in the VGG benchmark SYCL-DNN running on the GPU consistently
outperforms MKL-DNN\@.

\section{Conclusion}

The SYCL programming model allows us to write highly parametrized compute
kernels, which can be instantiated to best suit the targeted hardware and so
obtain good performance across different hardware.

Overall, we have shown that our general purpose, parameterized SYCL kernels give
competitive performance against hand tuned, optimized libraries such as clBLAST,
ARM Compute Library and MKL-DNN\@. Moreover these kernels can be easily tuned to
perform well on other devices, even if they have significantly different
performance characteristics.

There are many improvements still to make to these libraries, including adding
vector operations to SYCL-BLAS and implementations of convolution kernels in
SYCL-DNN which support data prefetching and local memory, as well as providing
currently unsupported operations required for neural networks. There are also
plans to develop a machine learning system to tune these libraries for new
devices.

%
\bibliographystyle{ACM-Reference-Format}
\bibliography{sc2019}

%

\end{document}